\documentclass[sigconf]{acmart}

\usepackage{amsmath,amssymb,amsfonts}
\usepackage{algorithmic}
\usepackage{graphicx}
\usepackage{subfig}
\usepackage{textcomp}
\usepackage{xcolor}
\usepackage{hyperref}
\usepackage{booktabs}
\usepackage{multirow}
\usepackage[linesnumbered,ruled,vlined]{algorithm2e}
\AtBeginDocument{%
  \providecommand\BibTeX{{%
    \normalfont B\kern-0.5em{\scshape i\kern-0.25em b}\kern-0.8em\TeX}}}

\setcopyright{acmlicensed}
\copyrightyear{2024}
\acmYear{2024}
\acmDOI{10.1145/3673038.3673151}

\acmConference[ICPP '24]{The 53rd International Conference on Parallel Processing}{August 12--15, 2024}{Gotland, Sweden}
\acmBooktitle{The 53rd International Conference on Parallel Processing (ICPP '24), August 12--15, 2024, Gotland, Sweden}
\acmPrice{}
\acmDOI{10.1145/3673038.3673151}
\acmISBN{979-8-4007-1793-2/24/08}
\begin{document}

\title{FedClust: Tackling Data Heterogeneity in Federated Learning through Weight-Driven Client Clustering}


\author{Md Sirajul Islam}
\affiliation{%
  \
  \institution{School of Computing and Informatics\\
  University of Louisiana at Lafayette}
  \streetaddress{1 Th{\o}rv{\"a}ld Circle}
  \city{Lafayette, LA}
  \country{USA}}
\email{md-sirajul.islam1@louisiana.edu}

\author{Simin Javaherian}
\affiliation{%
  \
  \institution{School of Computing and Informatics\\
  University of Louisiana at Lafayette}
  \streetaddress{1 Th{\o}rv{\"a}ld Circle}
  \city{Lafayette, LA}
  \country{USA}}
\email{simin.javaherian1@louisiana.edu}

\author{Fei Xu}
\affiliation{%
  \
  \institution{School of Computer Science and Technology\\
  East China Normal University}
  \streetaddress{1 Th{\o}rv{\"a}ld Circle}
  \city{Shanghai}
  \country{China}}
\email{fxu@cs.ecnu.edu.cn}

\author{Xu Yuan}
\affiliation{%
  \
  \institution{Department of Computer and Information Sciences\\
  University of Delaware}
  \streetaddress{1 Th{\o}rv{\"a}ld Circle}
  \city{Newark, DE}
  \country{USA}}
\email{xyuan@udel.edu}

\author{Li Chen}
\affiliation{%
  \
  \institution{School of Computing and Informatics\\
  University of Louisiana at Lafayette}
  \streetaddress{1 Th{\o}rv{\"a}ld Circle}
  \city{Lafayette, LA}
  \country{USA}}
\email{li.chen@louisiana.edu}

\author{Nian-Feng Tzeng}
\affiliation{%
  \
  \institution{School of Computing and Informatics\\
  University of Louisiana at Lafayette}
  \streetaddress{1 Th{\o}rv{\"a}ld Circle}
  \city{Lafayette, LA}
  \country{USA}}
\email{nianfeng.tzeng@louisiana.edu}

\renewcommand{\shortauthors}{Sirajul and Li, et al.}

\begin{abstract}
  Federated learning (FL) is an emerging distributed machine learning paradigm that enables collaborative training of machine learning models over decentralized devices without exposing their local data. One of the major challenges in FL is the presence of uneven data distributions across client devices, violating the well-known assumption of independent-and-identically-distributed (IID) training samples in conventional machine learning. To address the performance degradation issue incurred by such data heterogeneity, clustered federated learning (CFL) shows its promise by grouping clients into separate learning clusters based on the similarity of their local data distributions. 
  However, state-of-the-art CFL approaches require a large number of communication rounds to learn the distribution similarities during training until the formation of clusters is stabilized. Moreover, some of these algorithms heavily rely on a predefined number of clusters, thus limiting their flexibility and adaptability. In this paper, we propose {\em FedClust}, a novel approach for CFL that leverages the correlation between local model weights and the data distribution of clients. {\em FedClust} groups clients into clusters in a one-shot manner by measuring the similarity degrees among clients based on the strategically selected partial weights of locally trained models. We conduct extensive experiments on four benchmark datasets with different non-IID data settings. Experimental results demonstrate that {\em FedClust} achieves higher model accuracy up to $\sim$45\% as well as faster convergence with a significantly reduced communication cost up to 2.7$\times$ compared to its state-of-the-art counterparts.
\end{abstract}




\maketitle

\section{Introduction}
With the proliferation of Internet-of-Things (IoT) and the widespread adoption of artificial intelligence across various application domains,  machine learning has been increasingly shifted toward the network edge, where computations are performed on edge devices rather than in centralized data centers \cite{kairouz2021advances, mcmahan2017communication}. Such a computing paradigm shift is enabled by the rapid development of the computation and storage capacity on edge devices, able to handle more complex and data-intensive tasks. To analyze and process massive data generated by various edge devices ({\em e.g.}, mobile phones, wearable devices, and autonomous vehicles), the traditional machine learning approach falls short. It requires transmitting large volumes of user data to centralized cloud servers, incurring prohibitive communication costs and raising privacy concerns as well.

Federated learning (FL) has become a promising solution, allowing for participating devices to collaboratively train a globally shared model under the coordination of a central server, without exposing their local data. Due to its superior privacy-preservation implications, FL has been widely adopted by numerous companies (such as Google \cite{hard2018federated}) in a variety of applications, including computer vision \cite{liu2020fedvision}, natural language processing [23], and human activity recognition \cite{ouyang2021clusterfl, tu2021feddl}.
Essentially, FL is a distributed machine learning framework, with training data resident on decentralized client devices.
In classical FL training \cite{mcmahan2017communication, karimireddy2020scaffold, li20201federated}, the server broadcasts the current global model to the participating clients. Each client trains the model using its local data and sends its local updates to the server. The server then aggregates model updates from participating clients to update the global model, to be trained in the next round. These steps repeat until achieving a certain level of model accuracy or a pre-specified number of communication rounds. 

However, deploying FL often involves a number of devices that generate heterogeneous data due to their varying use styles. For instance, different users may watch videos on diverse types of content ({\em e.g.}, news, sports, and entertainment) and run their smartphones in varying frequencies. The presence of heterogeneous data across client devices breaks the conventional assumption of independent-and-identically-distributed (IID) training data, raising the new challenge of non-IID data distribution in the FL paradigm. 
Such data heterogeneity not only increases the overall communication cost but also degrades global model performance \cite{li2019convergence, zhao2018federated}, increasingly drawing research attention to mitigate the adverse impact of non-IID data on FL \cite{ghosh2020efficient, sattler2020clustered, duan2020fedgroup, briggs2020federated, ouyang2021clusterfl,kairouz2021advances}. 

Instead of learning a single global model, an alternative approach focuses on attaining personalized models for individual users to arrive at personalized FL, motivated by the observation that the globally learned model may exhibit lower accuracy on a participating device than a local model independently trained on its own data \cite{hanzely2020federated}.
More specifically, in addition to obtaining a collaboratively-trained global model, each client learns its personalized local model with various techniques such as regularization, local fine-tuning, model interpolation, multi-task learning, and knowledge distillation \cite{jiang2019improving, fallah2020personalized, mansour2020three, li2021ditto, li2019convergence, smith2017federated, luo2022pgfed, tan2022towards}.
Nevertheless, apart from a lack of generalization by its nature, personalized FL suffers from limited scalability, due to the extra computation overhead for learning a personalized model on each participant device. Additionally, it often fails to learn effectively on limited participant-specific data, unable to accurately capture local data distributions.

Recently, clustered federated learning (CFL) \cite{ghosh2020efficient, sattler2020clustered, ouyang2021clusterfl, islam2024fedclust} has gained significant attention as a promising solution for tackling data heterogeneity. CFL frameworks 
group clients into multiple clusters based on the similarity of their data distributions and train a separate model for each cluster to alleviate the adverse effect of non-IID data.  
Most of the existing CFL approaches indirectly measure the data distribution similarity among clients by utilizing their local model updates or the gradients ({\em e.g.}, \cite{sattler2020clustered}). 
Despite the promise of CFL, {\em it remains an open challenge for clustering clients optimally}. 
Existing efforts \cite{ghosh2020efficient, ouyang2021clusterfl, tu2021feddl} rely on a predefined cluster count,
which is difficult to determine optimally without any prior knowledge about data distributions or learning tasks among the clients. Under the assumption that the server holds a portion of globally shared data, Morafah \textit{et al.} \cite{morafah2023flis}  proposed to cluster clients based on the similarity of the inferences on the shared data using updated local models from clients. Such a data availability assumption on the server may not be practical in reality. Sattler \textit{et al.} \cite{sattler2020clustered} proposed to iteratively bi-partition clients into clusters based on the cosine similarity among their local model updates. This approach is communication-inefficient as it requires a large number of communication rounds to form stabilized clusters. Moreover, state-of-the-art CFL methods hardly allow the flexibility of balancing generalization and personalization.

To address the aforementioned limitations, we propose a novel clustered federated learning method, named {\em FedClust}, which efficiently groups clients with non-IID data into suitable clusters. The design of {\em FedClust} leverages our insight into the implicit relationship between the local model weights and the underlying data distribution on a client device. In particular, {\em FedClust} utilizes locally-trained model weights on a client, obtained by performing a few local training iterations on its own data. To measure the similarity among clients, {\em FedClust} requires each client to send only the strategically selected partial weights to the server, further reducing the amount of data transmission. After receiving weights from all the clients, {\em FedClust} constructs a proximity matrix based on the Euclidean distance to efficiently identify distribution similarities among clients. An agglomerative hierarchical clustering (bottom-up approach) \cite{day1984efficient} on the proximity matrix is employed to classify similar clients into an optimal number of clusters. Remarkably, {\em FedClust} operates without the need for any proxy data on the server, as opposed to all the existing solutions  \cite{morafah2023flis,gong2022adaptive}. 
Furthermore, in contrast to its state-of-the-art counterparts \cite{ghosh2020efficient, sattler2020clustered, briggs2020federated} which fail to accommodate client dynamics, 
{\em FedClust} provides an elegant mechanism for effectively incorporating new clients into appropriate clusters on-the-fly.

Finally, we have conducted extensive experiments to evaluate the performance of {\em FedClust} on four benchmark image classification datasets under different non-IID data settings, with our state-of-the-art counterparts following the LeNet-5 \cite{lecun1989backpropagation} and ResNet-9 \cite{he2016deep} architectures. The experimental results demonstrate that the proposed approach significantly improves overall model accuracy, surpassing both global and personalized baselines by up to $\sim$45\% and $\sim$18\%, respectively. Moreover, {\em FedClust} enables faster convergence due to great reduction in communication costs by 1.2 - 2.7$\times$  when compared to state-of-the-art FL methods. 

Our key contributions are summarized as follows:

\begin{itemize}
\item We analyze the relationship of model weights with the underlying data distribution of clients. Moreover, we observe the implicit connection of different layer weights with the local data distributions. 
  
\item We propose a novel clustered federated learning framework named {\em FedClust} to alleviate the adverse impact of non-IID data on FL. Our framework utilizes the similarity among strategically selected partial weights of locally-trained models from participant devices, to optimally form clusters for efficient and effective learning. In addition, we present an elegant strategy to accommodate client dynamics, incorporating newcomers into appropriate clusters in real time.


\item We conduct extensive experiments under diverse representative settings to evaluate {\em FedClust} with a variety of performance metrics. Experimental results demonstrate the advantages of {\em FedClust} over the state-of-the-art baselines, especially in improving overall model accuracy and communication efficiency.

\end{itemize}

The remainder of this paper is organized as follows. Section 2 describes related work. The background and motivation of this work are introduced in Section 3. In Section 4, we present a brief explanation on the design of our proposed {\em FedClust} framework. In Section 5, we evaluate {\em FedClust} and compare its performance results with those of state-of-the-art counterparts. Finally, Section 6 concludes the paper.

\section{Related Work}
In this section, we review pertinent work on addressing the non-IID data issue in federated learning.
\subsection{Federated Learning with Non-IID Data}
In federated learning (FL), data stored in each user device greatly varies due to different usage patterns and habits. Specifically, different users may prefer to browse news on different topics {\em e.g.}, sports, politics, and technology, leading to non-IID data for different users. In reality, it is natural that the data used for FL training are usually non-IID. The most widely used FL algorithm {\em FedAvg} \cite{mcmahan2017communication} fails to achieve optimal performance in the presence of non-IID data across clients. Several studies \cite{li20201federated, zhao2018federated} have shown that non-IID data not only decreases the accuracy of the trained model but also slows down training convergence with larger communication costs. To mitigate the client drift issue caused by non-IID data, {\em FedProx} \cite{li20201federated} introduces a proximal term to the local training objective to keep local models close to the global model. {\em FedDyn} \cite{acar2021federated} introduces a dynamic regularizer for each client in every round to align the global and the local models.

In {\em SCAFFOLD} \cite{karimireddy2020scaffold}, data heterogeneity is modeled as a source of variance among clients following a variance reduction technique. It estimates the direction of updates for the global model and that of each client. The drift of local training is then calculated by comparing two update directions. Finally, it modifies the local updates by incorporating the drift in local training. {\em FedNova} \cite{wang2020tackling} considers the number of local training epochs performed by each client during every round of FL to produce an unbiased global model. It normalizes and scales local updates based on the number of local training epochs before updating the global model. While effective under certain scenarios, these global FL methods cannot systematically address the data heterogeneity issue. Several studies \cite{lai2021oort, javaherian2024fedfair} proposed guided participant selection strategies, with a subset of clients is selected to participate in FL training according to some predefined criteria. These approaches provide faster model convergence and better time-to-accuracy performance. 

To mitigate the impact of non-IID data, prior approaches \cite{mansour2020three, jiang2019improving, fallah2020personalized} focused on personalizing the global model with each client's local data via fine-tuning. They let the global model act as an initial point for learning personalized models at clients based on their local data. Nevertheless, the global model may not be a good initiator, if the local data distributions of clients highly differ among one another. Similarly, Smith \textit{et al.} \cite{smith2017federated} extended multitask learning in FL training to aim at learning personalized models for multiple related tasks with the coordination of a central server. Recent work, {\em PGFed} \cite{luo2022pgfed}, formulated client's local objectives as personalized global objectives to explicitly transfer collaborative knowledge across them. 

\subsection{Clustered Federated Learning}
Alternatively, clustered federated learning (CFL) approaches \cite{ghosh2020efficient, sattler2020clustered, briggs2020federated, ouyang2021clusterfl} have been proposed to efficiently alleviate the challenge due to non-IID data among clients. They divide clients into clusters based on their data distributions so that clients in each cluster collaboratively train one model. Existing CFL approaches mainly differ in the process of identifying the data distribution similarity among clients by using model weights, gradient updates, or local loss.\;

Since the server cannot access clients' data in FL, \cite{sattler2020clustered, duan2020fedgroup} proposed to cluster clients based on the cosine similarity of their local model updates or gradients.  More specifically, {\em FMTL} \cite{sattler2020clustered} iteratively bi-partitions clients into clusters. Initially, all clients train a shared model by forming one cluster. Later, the server partitions the initial cluster into two new clusters according to the cosine similarity of their local model updates. The above steps repeat until no new cluster is generated. {\em IFCA} \cite{ghosh2020efficient} requires a predefined number of clusters. Each client needs to download available models of all clusters at each round and selects one that provides the highest local test accuracy. It incurs large communication costs due to the constant communication between the server and every client to form clusters. Moreover, the performance of {\em IFCA} heavily depends on the prior settings of the cluster count, which is difficult to identify without the knowledge of client local data distributions. Both of the above methods require large numbers of communication rounds to form stable clusters. 

Several early pursuits \cite{briggs2020federated, ouyang2021clusterfl} determine the number of clusters using the distances of their local model weights. Gong \textit{et al.} \cite{gong2022adaptive} proposed to pre-cluster clients based on certain statistics of each client's local data, and then adjust initial clustering by considering the model distance calculated among clients using partial weights. However, sharing statistics of the client's data with the server compromises the privacy promise of FL. Recently, Vahidian \textit{et al.} \cite{vahidian2022efficient} proposed {\em PACFL}, which identifies data distribution similarities among clients by using the principal angles over the client data subspaces. Before starting federation, it applies truncated Singular Value Decomposition (SVD) on each client's local data to obtain a small set of principal vectors which represent its underlying data distribution.

\begin{figure*}
\centering
\subfloat[Layer 1 (CL)]
{\includegraphics[height=4cm, width=0.25\textwidth]{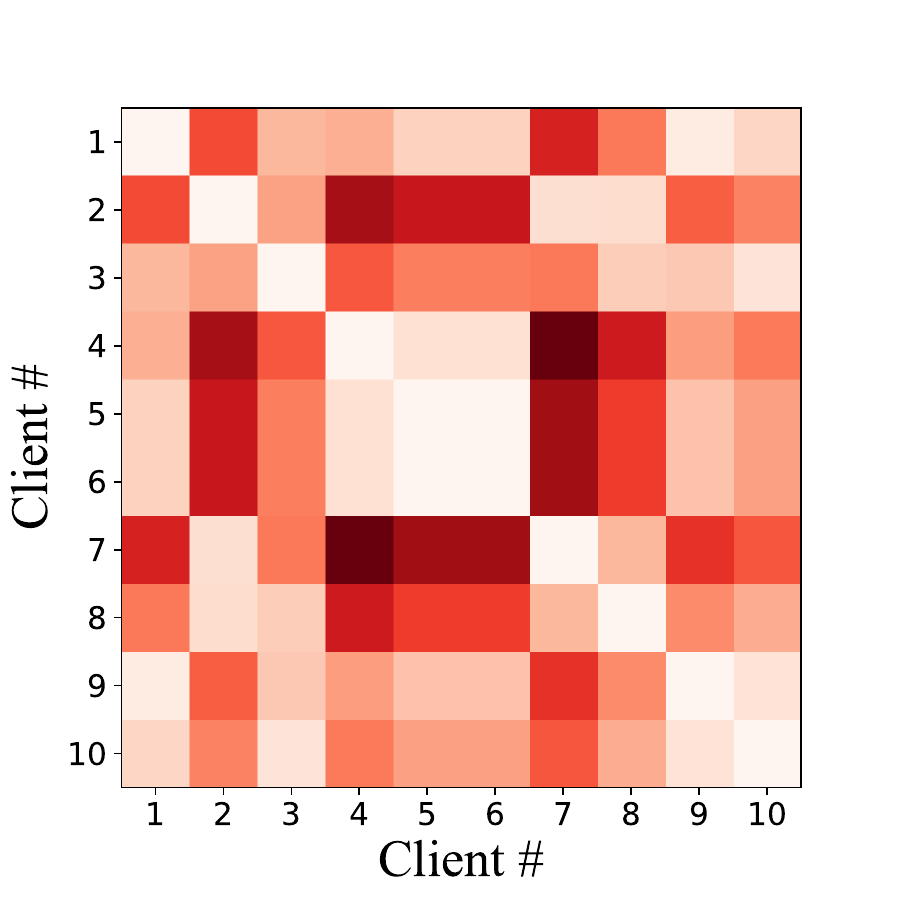}}
\subfloat[Layer 7 (CL)]
{\includegraphics[height=4cm, width=0.25\textwidth]{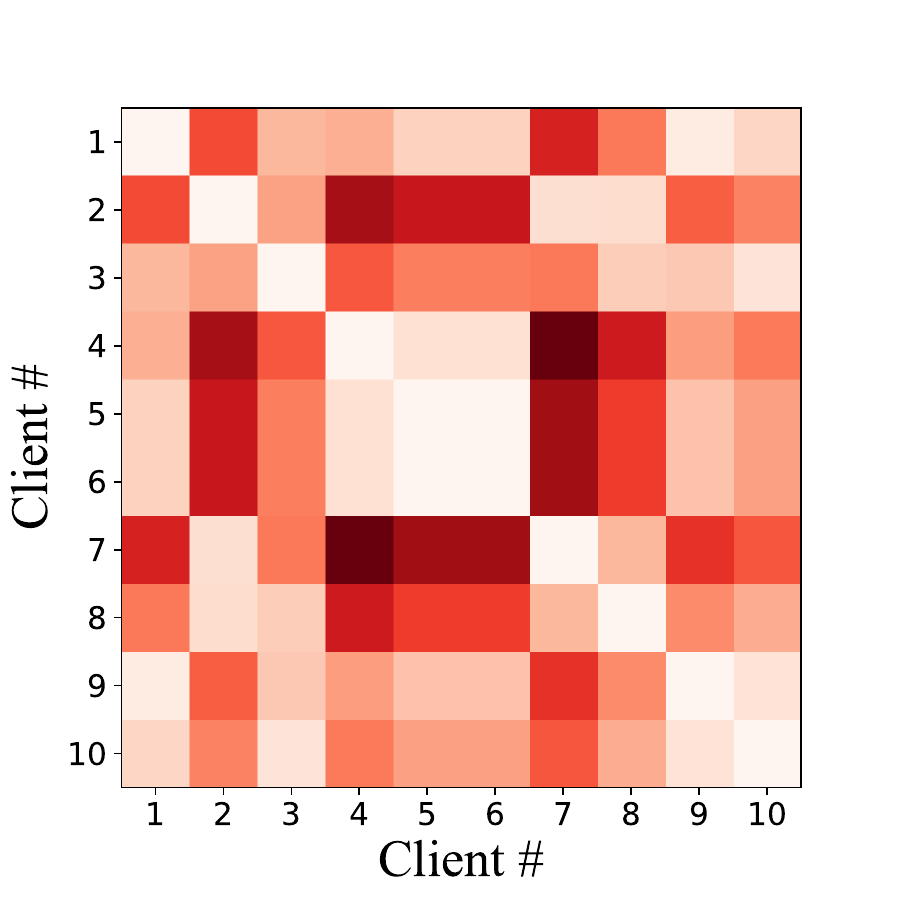}}
\subfloat[Layer 14 (FC)]
{\includegraphics[height=4cm, width=0.25\textwidth]{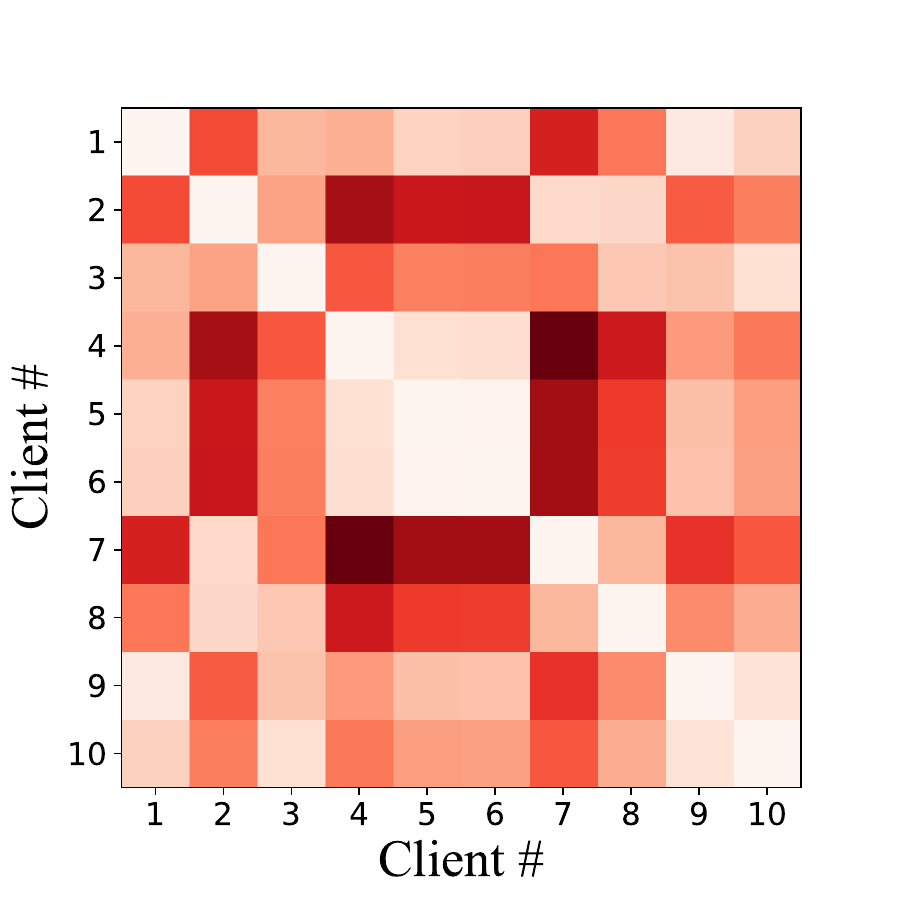}}
\subfloat[Layer 16 (FC)]
{\includegraphics[height=4cm, width=0.25\textwidth]{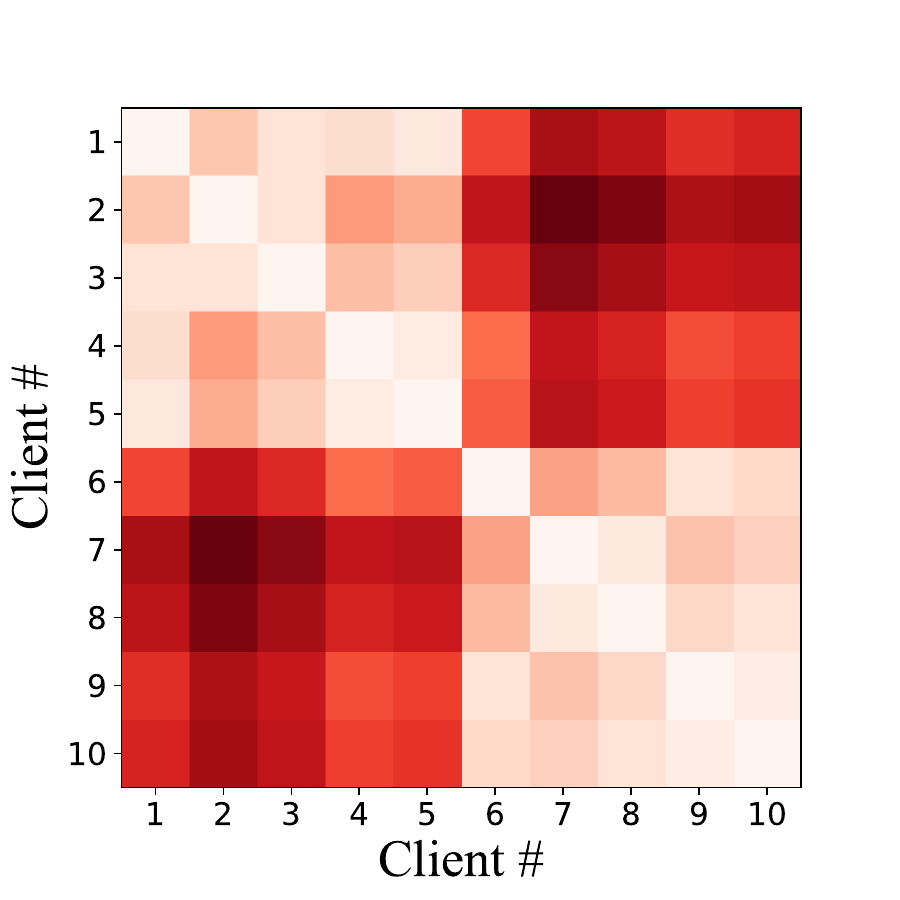}}\\
\caption{Illustration of the distance matrices calculated using different layer weights, where CL indicates convolutional layer and FC indicates fully connected layer. A lighter color in the distance matrices denotes a smaller distance, i.e., the two models are more similar. }
\label{figure}
\end{figure*}

\section{Background And Motivation}
\subsection{Federated Learning}
Federated learning (FL) is a privacy-preserving framework that allows distributed clients to collaboratively train machine learning or deep learning models without sharing their local data \cite{mcmahan2017communication}. FL usually involves a set of clients and a central server. Each clients receives its initial global model from the server and then train it for a few local iterations using its local data. The server is responsible for aggregating all local model updates to update the globally shared model. The communication between the server and every client follows a predetermined communication protocol. The server has no prior knowledge about the data distribution across devices as it cannot access the raw data stored in clients.

McMahan \textit{et al.} \cite{mcmahan2017communication} first introduced a federated averaging algorithm ({\em FedAvg}) that implements the idea of federated learning. In order to optimize the communication efficiency of FL over real-world data, the {\em FedAvg} algorithm trains a globally shared model across clients by a weighted averaging of the local model parameters of clients. In particular, the goal of {\em FedAvg} is typically to minimize the following objective function:
\begin{equation}
\min_{\theta} F(\theta) \overset{\scriptscriptstyle\Delta}{=} \sum_{i=1}^{m}\frac{n_i}{N}F_i(\theta)
\end{equation}
Here, $m$ is the set of participating clients and client $i$ has local dataset $\mathcal{D}_i$, where $n_i =|\mathcal{D}_i|$ and $N = \sum_{i=1}^{m} n_i$. The local objective functions of clients can be defined as the empirical loss over their local data $\mathcal{D}_i$, i.e., $F_i(\theta)=\frac{1}{n_i}\sum_{j_i=1}^{n_i}f_{j_i} (\theta; x_{j_i},y_{j_i})$, where $n_i$ is the number of client i's local samples. It is empirically shown that {\em FedAvg} provides better performance when the data distribution across clients is IID \cite{mcmahan2017communication}. In reality, data produced by different clients are usually non-IID in nature, thus negatively impacting the convergence and performance of federated learning in practical applications. Recently, clustered federated learning (CFL) has attracted research efforts to address the data heterogeneity issue with promising performance improvement.

\subsection{Motivation}
Although CFL-based approaches \cite{ghosh2020efficient, sattler2020clustered, briggs2020federated, ouyang2021clusterfl} have shown lofty improvement over {\em FedAvg} when dealing with non-IID data, they still lack efficiency due to their limitations of clustering strategies. We thus identify key limitations as well as opportunities in what follows:
\begin{itemize}
\item \textbf {Difficult to determine the cluster count in advance.} Most of the existing CFL approaches \cite{ghosh2020efficient, ouyang2021clusterfl, tu2021feddl} require a given number of clusters apriori, usually very hard to determine without knowing the actual data distributions across clients, despite that model accuracy is highly dependent on the optimal number of clusters.
\item \textbf {Require larger communication rounds to form stable clusters.} 
Some existing CFL approaches \cite{sattler2020clustered} can group clients into an appropriate number of clusters. Specifically, {\em FMTL} \cite{sattler2020clustered} iteratively partitions clients into clusters according to the cosine similarity of their local model updates. While yielding an optimal number of clusters, it is not communication efficient as a large number of communication rounds are needed to form stabilized clusters.
\item \textbf {Is it necessary to utilize all model weights?} A majority of current CFL approaches use all model weights or model updates to calculate model similarity which reflects the underlying data distribution of clients. It imposes a huge pressure on the server when calculating the similarity over a large number of models simultaneously. In addition, existing literature \cite{long2018transferable, yosinski2014transferable, rozantsev2018beyond} demonstrates that there are distinctions between the different layers in the same model, and higher layers weights are more task-related compared to lower layers weights. So, is it possible to effectively compare model similarity using just partial weights?
\end{itemize}
To address the above limitations, we propose a new clustered federated learning approach, {\em FedClust}, that divides clients into a suitable number of clusters in a one-shot manner by calculating the similarity using selected partial weights of clients' locally trained models.

\begin{figure*}
\centering
\includegraphics[height=7cm, width=0.8\textwidth]{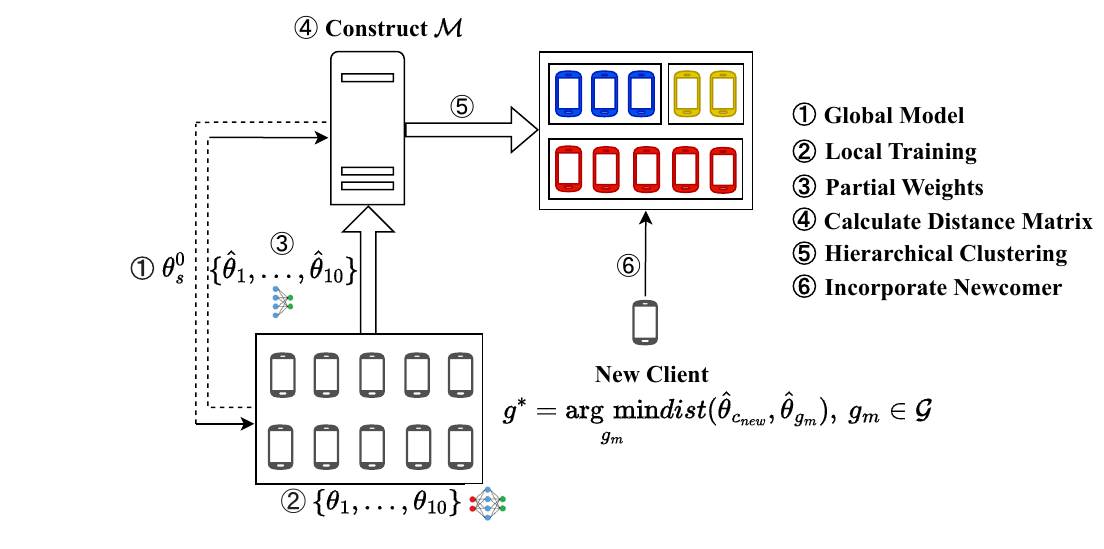}
\caption{An overview of FedClust. }
\label{figure}
\end{figure*}

\subsection{Observation}

Despite previous observations  \cite{long2018transferable, yosinski2014transferable, rozantsev2018beyond} on model layers as discussed before, there is a lack of understanding in how these variations can potentially impact federated learning. In this section, we conduct an experimental study to investigate the implications of model weights from different layers on the underlying data distribution.

We conduct a simple experiment for a multi-class image classification task on the CIFAR-10 \cite{krizhevsky2009learning} dataset with VGG16 \cite{simonyan2014very}. Specifically, the VGG16 model contains thirteen convolutional layers and three fully connected layers. To simulate non-IID data among clients, we assume 10 different clients and group them into two clusters based on their local label categories, {\em e.g.}, $\mathcal{G}_1 = \{1, 2, \ldots, 5\}$ and $\mathcal{G}_2 = \{6, 7, \ldots, 10\}.$ Fig. 1 illustrates four different distance matrices which are calculated using the weights of the four different layers (1, 7, 14, and 16) in VGG16.

From Fig. 1, we observe that the final layer weights implicitly represent the underlying data distribution of clients. Specifically, Figs. 1(a) and 1(b) depict the distance matrices based on the two convolutional layer weights respectively. However, we cannot obtain the cluster structures of the clients from them. The clustering structures of the clients are clearly observed in Figs. 1(d). Moreover, we observed similar clustering patterns using the LeNet-5 model. Based on the above experiments and previous studies \cite{long2018transferable, yosinski2014transferable, rozantsev2018beyond}, we can conclude that the final layer or the layer with the classifier function reflects the model difference caused by non-IID data. In addition, clients with similar data distributions tend to train the model in a similar manner, resulting in closer distances among final layer'r weights.

\subsection{Overview}
To advance existing CFL pursuits, we propose {\em FedClust} which can identify cluster patterns among clients based on the distance of weights from the final layers of their local models. The proposed method, as described in Algorithm 1, is able to alleviate the effects of non-IID data on practical FL applications. We first focus on clustering clients based on their local data distributions in a federated network. The proposed approach is one-shot clustering that classifies \textit{m} clients into \textit{n} clusters, i.e., $\mathcal{G} = \{g_1,\ldots, g_n\}$, based on the similarity of their underlying data distributions before starting federation. {\em FedClust} trains one model individually for each cluster $g_k$, instead of training a common global model for all clients. The objective function for the clients of each cluster ${g_k}$ is defined as follows:
\begin{equation}
\min_{\theta} F(\theta_{g_k}) \overset{\scriptscriptstyle\Delta}{=}\sum_{{c_i}\in{g_k}}{\frac{n_{c_i}}{N_{g_k}}F_{ik}(\theta_{g_k})}
\end{equation}
Here, $n_{c_i}$ and $N_{g_k}$ denotes the number of data samples for client $c_i$ and cluster $g_k$, respectively, and $F_{ik}(\theta_{g_k})$ determines the empirical loss on $c_i$'s local dataset  $\mathcal{D}_i$.

An overview of the proposed framework is depicted in Fig. 2. To minimize computation overhead, our clustering strategy employs a static approach that avoids rescheduling clients in each round. First, the server broadcasts the initial global model to all available clients. Then, each client trains the model on its local data for a few local iterations and sends back the updated final layer weights to the server as the representation of their underlying data distribution. The server then computes the proximity matrix between models based on the final layer weights uploaded by each client. Finally, the server employs agglomerative hierarchical clustering (HC) \cite{day1984efficient} on the proximity matrix $\mathcal{M}$ to group clients with similar data distribution into the same cluster. The basic idea is that, initially considering each client as a separate cluster, it repeatedly performs the following operations in each iteration: (1) identify the two clusters that exhibit the closest similarity, and (2) combine the two most similar clusters. To determine which clusters should be merged, a linkage criterion ({\em e.g.} single, average, complete, {\em etc.}) is defined \cite{day1984efficient}. For instance, the smallest distance between two points in each cluster in ``single linkage" is defined as the pairwise Euclidean distance between two clusters. In this paper, we denote $\lambda$ as the clustering threshold, representing the distance between two clusters. The iterative process continues until a suitable number of clusters have been formed. The above clustering process is done in one communication round.

From the next round, the workflow of {\em FedClust} is similar to {\em FedAvg} \cite{mcmahan2017communication}. The server initializes all cluster models with $\theta^{0}_{s}$. A subset of available clients is selected randomly by the server and the server broadcasts $\theta^{0}_{s}$ to selected clients. Each client trains the model on its local data and performs a few steps of stochastic gradient descent (SGD) updates. The clients send back their updated model parameters along with their cluster IDs to the server. The server conducts model averaging for each cluster after receiving model updates from all clients. 

\section{Methodology}
\SetKwInput{KwInput}{Input}              
\SetKwInput{KwOutput}{Init}           

\begin{algorithm}[t]
\DontPrintSemicolon
  \KwInput{Number of available clients $N$, number of communication rounds $T$, clustering threshold $\lambda$, sampling rate $R \in (0,1]$}
  \KwOutput{Server model initialization with $\theta^{0}_{s}$}
  \For{each round i = 0, 1, 2,...}{
        \uIf{i = 0}
          {
            Server broadcast $\theta^{0}_{s}$ to all available clients.\;
            Each client performs local update and sends back the updated weights of the last layer to the server.\;
            $\mathcal{M}\leftarrow$ server constructs ${\mathcal{M}}$ based on Eq. 3\;
            $\left\{C_1,\ldots,C_m\right\} = HC (\mathcal{M}, \lambda)$\ \tcp*{employing hierarchical clustering to obtain the clusters}
            $\theta^{0}_{s,m}\leftarrow\theta^{0}_{s}$\ \tcp*{clusters model initialization with $\theta^{0}_{s}$}   
            }
        \Else
        {
        $n\leftarrow max(R \times N, 1)$\;
        $\mathcal{S}_{i}\leftarrow \left\{k_1,...,k_n\right\}$ random set of n clients\;
        }\
        \For{each client $k \in \mathcal{S}_{i}$ in parallel}
        {
        Client $k$ sends its cluster ID to the server and receives the corresponding cluster model from the server $\theta^{i}_{s,m}$\;
        ${\theta^{i+1}_{k,m}\leftarrow}$ ClientUpdate ${(\textit{k}; \theta^{i}_{s,m})}$\ \tcp*{local SGD training}}\
        $\theta^{i+1}_{s,m} = \sum_{k\in{C_m}}|D_k|\theta^{i+1}_{k,m} /\sum_{k\in{C_m}}|D_k|$\ \tcp*{each cluster model averaging}  
        }
    
\caption{FedClust}
\end{algorithm}

\subsection{Selection of Model Weights}
{\em FedClust} utilizes only the final layer weights instead of the full model weights to group clients into clusters. We have observed empirically that there exists an implicit relationship between the local data distribution of clients with the model weights trained on their dataset, consistent with the findings in \cite{wang2020optimizing}. Therefore, {\em FedClust} leverages clients model weights to infer the relative characteristics of the underlying data distribution. The difference in data distribution between clients can be approximated based on the difference in their model weights, referred to as the model distance, after the completion of their local model training.  The model distance between the model weights of any two clients $c_p$ and $c_q$ can be calculated using $l_2$ distance as follows: 
\begin{equation}
dist(c_p, c_q)= \Big|\Big|\hat{\theta}_{c_p}-\hat{\theta}_{c_q}\Big|\Big|_{l_2}
\end{equation}

Generally, if two clients contain similar data distributions, they tend to train models in a similar fashion compared to clients with dissimilar data \cite{sattler2020clustered, ouyang2021clusterfl, wang2020optimizing}. As a result, the distance between their model weights will be smaller. Therefore, the model distance can be used as a useful metric to cluster clients. The server forms a distance matrix $\mathcal{M}$ of size $m\times m$ after receiving the final layer weights of all client's models. Each entity of the matrix $\mathcal{M}_{pq}$ represents the computed model distance $dist(c_p, c_q)$ between clients $c_p$ and $c_q$. Federated learning training typically involves a large number of client devices and the target machine learning model could be complex with huge parameters, {\em e.g.}, the VGG16 model has a total of 138M weights \cite{simonyan2014very}. Consequently, clustering methods based on model weights would require higher computation costs, affecting the clustering efficiency. 

To enhance clustering accuracy and reduce additional computational overheads, we only select the final layer weights of the clients local model instead of full model weights to determine the similarity. In Fig. 1, we have empirically demonstrated that the final layer weight of the model reflects the model difference caused by non-IID data. Therefore, calculating the distance matrix using all weights can lead to a bad similarity matrix thus reducing the clustering accuracy.
Moreover, the lower layers of the model contain the majority portion of the weights. Specifically, in deep learning models, especially those used for image classification tasks, {\em e.g.}, CNN models, the purpose of convolutional layers is to identify and extract features from the input, while the fully connected layers focus on the final classification task. Therefore, fully connected layer weights are more task-related. In {\em FedClust}, we thus choose a subset of the model's parameters, specifically the weights and bias of the final layer, to serve as a representation of the entire model. We utilize these weights to calculate distance matrix $\mathcal{M}$. It significantly reduces the computation cost as the size of the final layer weights ${\hat{\theta}_{c_p}}$ is much smaller than the full model weights ${{\theta}_{c_p}}$.

\begin{algorithm}[t]
\DontPrintSemicolon
  \KwInput{New client \textit{$c_{new}$}, Existing clusters partial model weights $\{\hat{\theta}_{g_{1}},\ldots,\hat{\theta}_{g_{m}}\}$}
  \KwOutput{Appropriate cluster $g^*$ to incorporate $c_{new}$}
  Server sends the initial global model $\theta^{0}_{s}$ to $c_{new}$\;
  $\theta_{c_{new}}\leftarrow$ $c_{new}$ trains the model using its local data\;
  ${\hat{\theta}_{c_{new}}}\leftarrow$ Transmit partial model weights to the server\;
  Server assigns the $c_{new}$ to cluster $g_m$:\;
  ${g^*}= \operatorname*{arg\,min}_{g_m}dist({\hat{\theta}_{c_{new}}}, {\hat{\theta}_{g_{m}})},\:  {g_m \in {\mathcal{G}}}$\
\caption{Incorporating Newcomers}
\end{algorithm}

\subsection{Incorporating Newcomers}
In reality, client devices may join in or drop out of the federated learning process due to unreliable client communications or other resource limitations. Clients who quit the training have no impact on the model training of their respective clusters. However, it is important to carefully incorporate newcomer clients into appropriate clusters in order to maintain the scalability of our method. {\em FedClust} offers an elegant approach to accommodate newcomers who join after the federation procedure to learn their personalized model. The baseline methods except {\em PACFL} \cite{vahidian2022efficient} did not clarify the process of incorporating newcomer clients during federation. We outlined the process of how {\em FedClust} integrates new participants who joined after the end of the federation in Algorithm 2. 

Clients who are not in the existing client set are referred to as newcomers. In order to assign each new client $c_{new}$ into a suitable cluster, $c_{new}$ is required to train the initial server model $\theta^{0}_{s}$ on its local data, and then sends partially selected weights to the server. {\em FedClust} stores a copy of each cluster's partial model weights. After receiving the partial model weights from new client $c_{new}$, the server computes the model distances between the new client $c_{new}$'s model and the models of existing clusters. The cluster with the minimum model distance will be selected as the cluster for $c_{new}$, represented as follows.
\begin{equation}
{g^*}= \operatorname*{arg\,min}_{g_m}dist({\hat{\theta}_{c_{new}}}, {\hat{\theta}_{g_{m})},\:  {g_m \in {\mathcal{G}}}}
\end{equation}
\SetKwInput{KwInput}{Input}              
\SetKwInput{KwOutput}{Output}           

\section{Experiments}
\subsection{Experimental Setup}

\textbf{Datasets and Models.} 
We evaluate the performance of {\em FedClust} on different image classification tasks using four popular benchmark datasets, i.e., CIFAR-10 \cite{krizhevsky2009learning}, CIFAR-100 \cite{krizhevsky2009learning}, Fashion MNIST (FMNIST) \cite{xiao2017fashion}, and SVHN \cite{netzer2011reading}. To imitate non-IID scenarios, we consider three different data heterogeneity settings for each dataset as in \cite{li2022federated}, i.e.,  Non-IID label skew (20\%), Non-IID label skew (30\%), and Non-IID Dir (0.1). In our experiments, we consider LeNet-5 \cite{lecun1989backpropagation} architecture for CIFAR-10, FMNIST, and SVHN datasets and ResNet-9 \cite{he2016deep} architecture for CIFAR-100 dataset. 

\textbf{Baselines Methods.} To demonstrate the performance of the proposed method, we compare the results of {\em FedClust} against the following state-of-the-art (SOTA) FL baselines. We consider {\em FedAvg} \cite{mcmahan2017communication}, {\em FedNova} \cite{wang2020tackling}, and {\em FedProx} \cite{li20201federated} for baselines that train a single global model across all clients. Baselines for SOTA CFL methods include {\em IFCA} \cite{ghosh2020efficient}, {\em PACFL} \cite{vahidian2022efficient} and Clustered-FL ({\em CFL}) \cite{sattler2020clustered}. SOTA personalized FL methods include {\em Per-FedAvg} \cite{fallah2020personalized} and {\em LG-FedAvg} \cite{liang2020think}. In addition, we compare our results with another baseline named {\em Local}, where each client independently trains a model on its local data without any communication with others.

\textbf{Implementation.} We have implemented {\em FedClust} and the baseline methods in PyTorch \cite{paszke2019pytorch}. We assume 100 clients are available for all experiments and 10\% of them are sampled randomly in each communication round.  We ran each experiment 3 times for 200 communication rounds. We execute all experiments on a server, which is equipped with NVIDIA GeForce RTX 3080Ti GPU, Intel(R) Core(TM) i9-10900X CPU, and 64G RAM. We emulate both the server and clients on the same machine, substantiated by the fact that the performance metrics we consider remain unaffected by the physical location of the server and clients. The wall-clock training time may be affected but this metric is beyond our scope of focus, similar to our counterparts.

\textbf{Hyperparameters Settings.}
In all of our experiments, we use SGD as the local optimizer with the local epoch of 10, and the local batch size of 10. We initialize the models randomly in {\em LG-FedAvg} for a fair comparison instead of using the model produced after many rounds of {\em FedAvg}. For {\em IFCA} and {\em CFL}, we used the same number of clusters as mentioned in the original papers. For {\em PACFL}, we used p = 3 in all of our experiments. The learning rate for {\em FedAvg}, {\em FedProx}, {\em FedNova}, and {\em CFL} was set to (0.1, 0.01, 0.001), while for other baselines, it was 0.01. Momentum was 0.9 for {\em FedAvg}, {\em FedProx}, and {\em FedNova}, whereas for other methods, it was 0.5. In {\em LG}, the number of local layers and global layers were set to 3 and 2. In {\em Per-FedAvg}, we used $\alpha=1e-2$ and $\beta=1e-3$. For {\em CFL}, the values of $\epsilon_1$ and $\epsilon_2$ were 0.4 and 0.6, respectively.

\textbf{Evaluation Metrics.} We use the average of the final local test accuracy over all clients and the number of communication rounds required to reach a certain level of model accuracy as the performance metrics. In general, it is desirable to achieve higher model accuracy with fewer communication rounds. We also consider the required communication costs to reach a target accuracy.

\begin{table}[t]
\caption{Test accuracy comparisons of different approaches over different datasets for Non-IID label skew of 20\%}
\centering
\large
\resizebox{\columnwidth}{!}{\begin{tabular}{lllll}
\toprule
\text{Method} & \text{CIFAR-10} & \text{CIFAR-100} &\text{FMNIST} &\text{SVHN}   \\
\midrule
Local&79.68 ± 1.32 &33.18 ± 0.41 & 95.68 ± 0.84 & 80.29 ± 1.61 \\
FedAvg&50.27 ± 2.63 &53.67 ± 0.63 & 77.10 ± 3.29 & 81.36 ± 0.64 \\
FedProx&51.60 ± 1.40 &54.28 ± 0.76 & 74.53 ± 2.16 & 79.64 ± 0.80 \\
FedNova&47.38 ± 2.08 &53.90 ± 0.38 & 71.33 ± 4.50 & 75.56 ± 3.07 \\
LG&85.49 ± 0.87 &54.15 ± 0.29 & 95.49 ± 0.75 & 91.59 ± 0.42 \\
PerFedAvg&85.80 ± 0.58 &61.29 ± 0.42 & 95.78 ± 1.28 & 92.87 ± 1.92 \\
CFL&51.86 ± 1.31 &41.28 ± 1.75 & 78.44 ± 2.38 & 73.59 ± 1.86 \\
IFCA&87.19 ± 0.19 &70.35 ± 0.28 & 96.83 ± 0.24 & 94.76 ± 0.19 \\
PACFL&88.40 ± 0.48 &71.06 ± 0.39 & 97.46 ± 0.13 & 95.48 ± 0.27 \\
\textbf {FedClust}&\textbf{95.82 ± 0.17} & \textbf{73.38 ± 0.24} & \textbf{97.92 ± 0.18} & \textbf{95.86 ± 0.11} \\
\bottomrule
\end{tabular}}
\end{table}
\begin{table}[t]
\caption{Test accuracy comparisons of different approaches over different datasets for Non-IID label skew of 30\%}
\centering
\large
\resizebox{\columnwidth}{!}{\begin{tabular}{lllll}
\toprule
\text{Method} & \text{CIFAR-10} & \text{CIFAR-100} &\text{FMNIST} &\text{SVHN}   \\
\midrule
Local&66.51 ± 0.92 & 23.76 ± 0.84 & 92.51 ± 0.24 & 68.84 ± 2.86 \\
FedAvg&57.79 ± 1.08 & 54.79 ± 0.56 & 79.90 ± 1.81 & 82.58 ± 0.75 \\
FedProx&56.92 ± 1.26 & 53.65 ± 0.60 & 81.53 ± 1.48 & 82.91 ± 1.30 \\
FedNova&54.15 ± 1.31 & 54.11 ± 0.95 & 78.02 ± 2.08 & 80.26 ± 1.49 \\
LG&75.42 ± 0.41 & 36.78 ± 0.68 & 94.54 ± 0.48 & 88.07 ± 0.65 \\
PerFedAvg&78.67 ± 0.32 & 57.02 ± 0.49 & 92.35 ± 1.70 & 92.10 ± 1.27 \\
CFL&52.03 ± 2.84 & 35.73 ± 2.14 & 78.38 ± 0.42 & 74.02 ± 3.90 \\
IFCA&80.21 ± 0.16 & 66.21 ± 0.21 & 95.29 ± 0.19 & 92.87 ± 0.14 \\
PACFL&82.35 ± 0.27 & 65.91 ± 0.17 & 95.43 ± 0.07 & 93.05 ± 0.18 \\
\textbf {FedClust}&\textbf{83.21 ± 0.25} & \textbf{68.33 ± 0.19} & \textbf{95.70 ± 0.09} & \textbf{93.17 ± 0.04} \\
\bottomrule
\end{tabular}}
\end{table}

\subsection{Results and Analysis}
\begin{figure*}
\centering
{\includegraphics[height=3.25cm, width=0.23\textwidth]{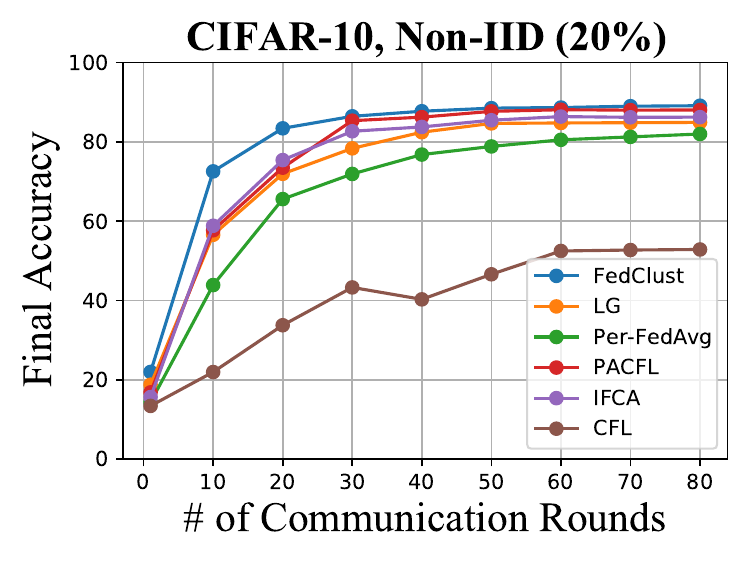}}
{\includegraphics[height=3.25cm, width=0.23\textwidth]{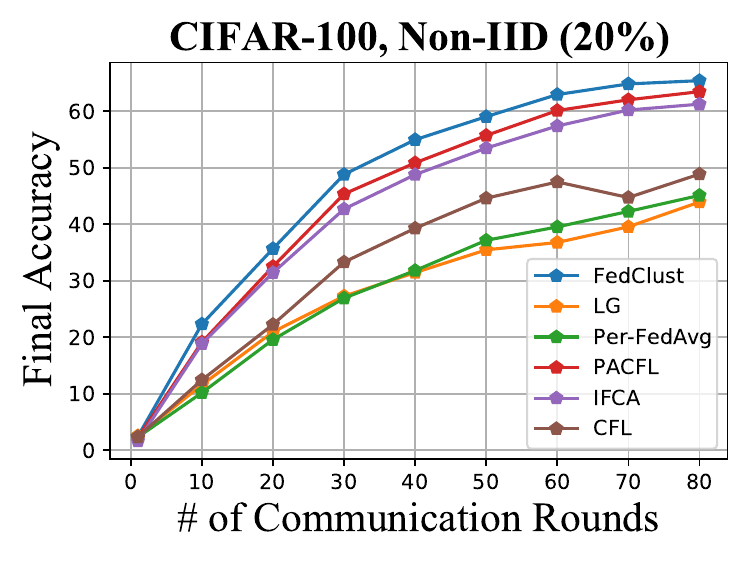}}
{\includegraphics[height=3.25cm, width=0.23\textwidth]{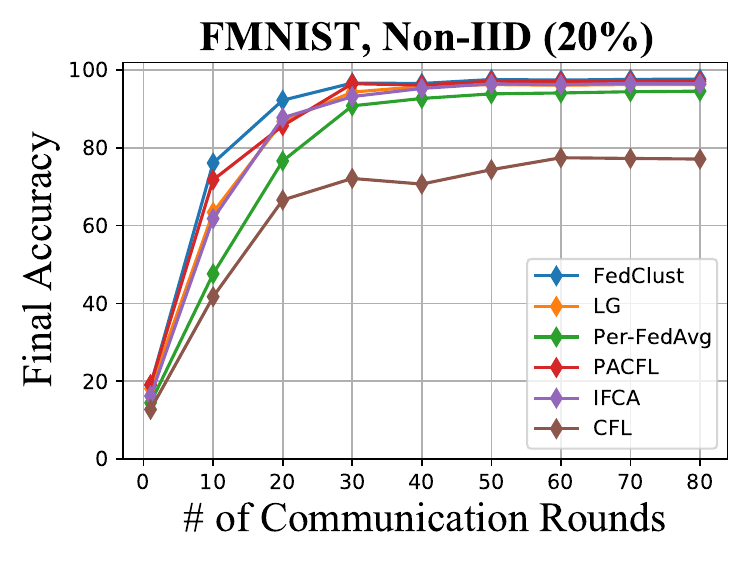}}
{\includegraphics[height=3.25cm, width=0.23\textwidth]{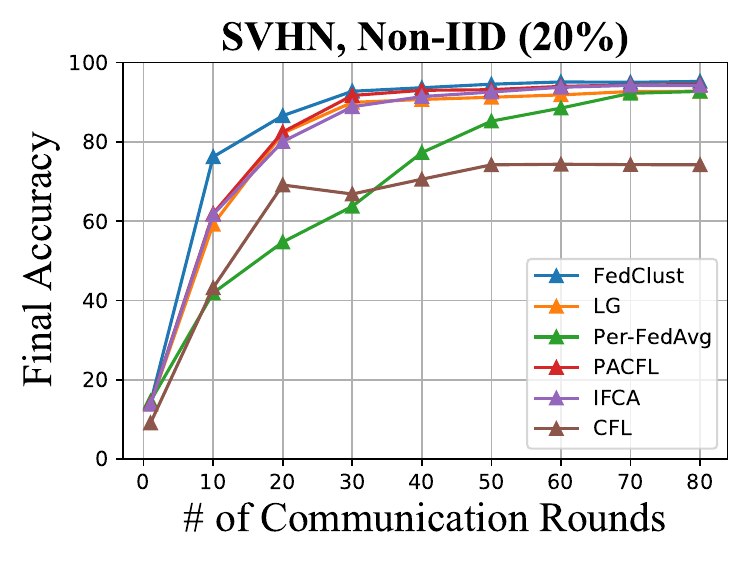}}
\caption{Test accuracy versus the number of communication rounds for Non-IID label skew of 20\%. FedClust converges faster to reach target accuracy and consistently outperforms other baselines. }
\label{figure}
\end{figure*}
\textbf{Performance comparisons.}
We compare {\em FedClust} with other SOTA baseline methods under two different widely used Non-IID settings, i.e. Non-IID label skew, and Non-IID Dirichlet label skew \cite{li2022federated}. We consider two different Non-IID label skew settings 20\% and 30\%. In both settings, we start by randomly assigning $\delta$\% of the total available labels of a dataset to each client and then randomly distributing the samples of each label among clients who own these labels. In Non-IID Dirichlet label skew settings, we assign training data to clients according to the Dirichlet distribution similar to \cite{li2022federated}. We run each experiment three times for 200 communication rounds with a local epoch of 10 and report the mean and standard deviation of the average of final local test accuracy.

Table 1, 2, and 3 show the performance comparisons among all approaches under Non-IID label skew (20\%), Non-IID label skew (30\%), and Non-IID Dir (0.1) settings, respectively. The results show that global FL baselines, i.e. {\em FedAvg}, {\em FedProx}, and {\em FedNova} provide poor performance in all scenarios due to the model drift and weight divergence issues. {\em Local} without any communications costs even perform better than global FL baselines. However, individual clients may not have enough data and thus clustering could be a promising solution. {\em FedClust} consistently demonstrates superior performance compared to SOTA baselines on all datasets for different data heterogeneity settings. In particular, focusing on the CIFAR-10 Non-IID label skew (20\%) setup, {\em FedClust} outperforms all SOTA global baseline methods (by +45.5\%, +44\%, +48.5\% for {\em FedAvg}, {\em FedProx}, {\em FedNova}) as well as personalized approaches (by +44\%, +10\%, +10\%, +8.5\%, +7\% for {\em CFL}, {\em LG}, {\em PerFedAvg}, {\em IFCA}, {\em PACFL}). We tuned the hyperparameters for each baseline in order to achieve the optimal outcome.

\begin{table}[t]
\caption{Test accuracy comparisons of different approaches over different datasets for Non-IID Dir (0.1)}
\centering
\large
\resizebox{\columnwidth}{!}{\begin{tabular}{lllll}
\toprule
\text{Method} & \text{CIFAR-10} & \text{CIFAR-100} &\text{FMNIST} &\text{SVHN}   \\
\midrule
Local&41.80 ± 2.09 & 17.56 ± 0.61 & 70.40 ± 0.86 &  59.06 ± 1.43  \\
FedAvg&38.25 ± 2.98 & 45.26 ± 0.47  & 81.93 ± 0.64 &  61.26 ± 0.95 \\
FedProx&42.69 ± 1.47 & 46.17 ± 0.83 & 83.32 ± 1.07 &  62.31 ± 1.72  \\
FedNova&39.52 ± 1.35 & 46.55 ± 1.54 & 83.68 ± 1.61 &  60.53 ± 2.18  \\
LG&48.63 ± 0.42 & 24.27 ± 0.33 & 74.39 ± 1.26 &  73.12 ± 0.76 \\
PerFedAvg&52.83 ± 1.47 & 34.20 ± 0.29 & 81.18 ± 1.80 & 75.07 ± 1.85 \\
CFL&41.50 ± 0.35 & 31.62 ± 1.76 & 74.01 ± 1.19 &  61.96 ± 1.58  \\
IFCA&50.51 ± 0.61 & 46.28 ± 0.23 & 84.57 ± 0.41 & 74.57 ± 0.40 \\
PACFL&51.02 ± 0.24 & 47.58 ± 0.20 & 85.30 ± 0.28 & 76.35 ± 0.46 \\
\textbf {FedClust}&\textbf{60.25 ± 0.58} &\textbf{49.65 ± 0.17} & \textbf{95.51 ± 0.17} & \textbf{78.23 ± 0.30} \\
\bottomrule
\end{tabular}}
\end{table}

\textbf{Communication cost.}
In this experiment, we compare the performance of our {\em FedClust} with baseline methods where the number of communication rounds for the federation is limited under a heterogeneous setting. Herein, we limit the communication round budget to 80 rounds for all personalized baselines. We illustrate the number of communication rounds versus the average of the final local test accuracy across all clients for Non-IID label skew (20\%) in Fig. 3. Our proposed approach achieves convergence within only 20 communication rounds in CIFAR-10, FMNIST, and SVHN datasets. From Fig. 3, we can see that the {\em CFL} baseline \cite{sattler2020clustered} shows the worst performance on all datasets, except for CIFAR-100. {\em Per-FedAvg} experiences greater advantages as the number of communication rounds increases. It appears that {\em PACFL} and {\em IFCA} are the closest lines to ours for all datasets, with {\em FedClust} consistently outperforming. The reason behind this fact is that {\em IFCA} requires many rounds of federation to form stabilized clusters as it initially starts with random cluster models which are inherently noisy.

\begin{table}[t]
\caption{Comparison of the number of communications rounds needed for different approaches to reach target top-1 average local test accuracy over different datasets for Non-IID of 20\%}
\centering
\resizebox{0.88\columnwidth}{!}{\begin{tabular}{lllll}
\toprule
\text{Method} & \text{CIFAR-10} & \text{CIFAR-100} &\text{FMNIST} &\text{SVHN} \\
\midrule
\text{Target} & \text{80\%} &\text{50\%} &\text{75\%} &\text{75\%} \\
\midrule
FedAvg&-- -- & 135 & 200 & 150 \\
FedProx&-- -- & 120 & 200 & 200 \\
FedNova&-- -- & 125 & -- -- & 150\\
LG&27 & -- -- & 14 & 17 \\
PerFedAvg& 54 & 110 & 15 & 37 \\
CFL&-- -- & -- -- & 47 & -- -- \\
IFCA&28 & 43 & 13 & 19 \\
PACFL&25 & 40 & 13 & 15 \\
\textbf {FedClust}& \textbf{13} & \textbf{32} & \textbf{7} & \textbf{9}\\
\bottomrule
\end{tabular}}
\end{table}

\begin{table}[t]
\caption{Comparison of the required communications costs in Mb needed for different approaches to reach target top-1 average local test accuracy over different datasets for Non-IID label skew of 30\%}
\centering
\resizebox{0.88\columnwidth}{!}{\begin{tabular}{lllll}
\toprule
\text{Method} & \text{CIFAR-10} & \text{CIFAR-100} &\text{FMNIST} &\text{SVHN}    \\
\midrule
\text{Target} & \text{70\%} &\text{50\%} &\text{80\%} &\text{80\%} \\
\midrule
FedAvg&-- -- & 4237.37 & 79.36 & 71.43 \\
FedProx&-- -- & 4237.37 & 71.43 & 71.43 \\
FedNova&-- -- & 3601.98 & -- -- & 79.36 \\
LG&\textbf{2.11} & -- -- & \textbf{1.26} & \textbf{1.76} \\
PerFedAvg& 23.81 & 6356.06 & 7.54 & 18.65 \\
CFL&-- -- & -- -- & -- -- & -- --\\
IFCA&16.66 & 3495.19 & 11.30 & 10.71 \\
PACFL&10.31 & 1991.60 & 7.53 & 8.73 \\
\textbf{FedClust}&8.66 & \textbf{1889.17} & 4.60 & 7.11 \\
\bottomrule
\end{tabular}}
\end{table}

\begin{figure*}
\centering
{\includegraphics[height=3.25cm, width=0.23\textwidth]{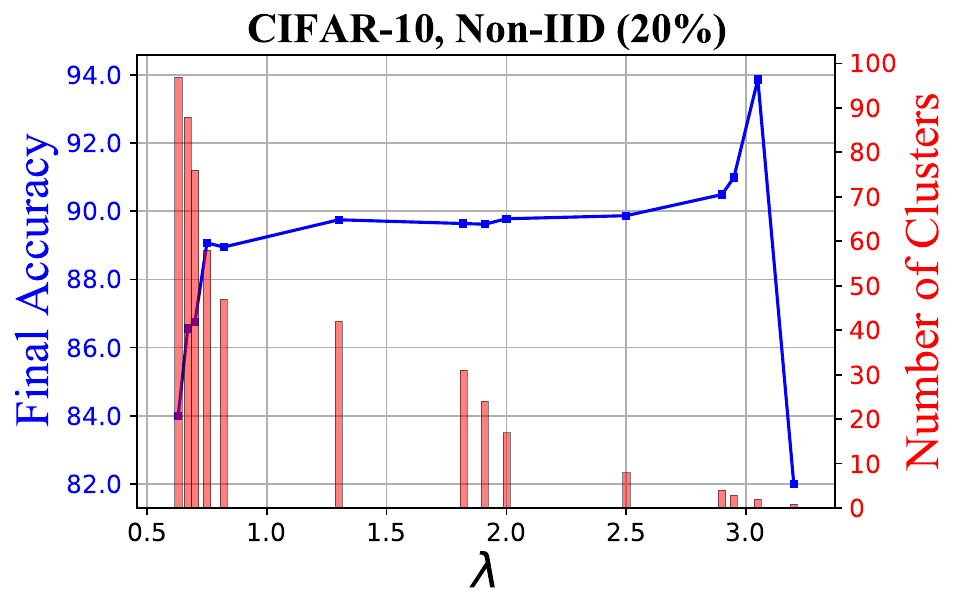}}
{\includegraphics[height=3.25cm, width=0.23\textwidth]{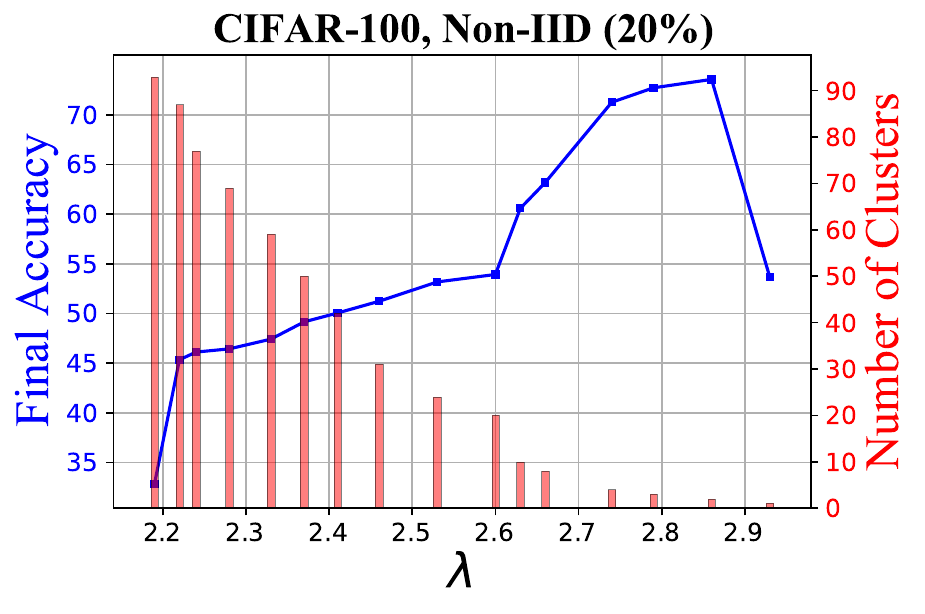}}
{\includegraphics[height=3.25cm, width=0.23\textwidth]{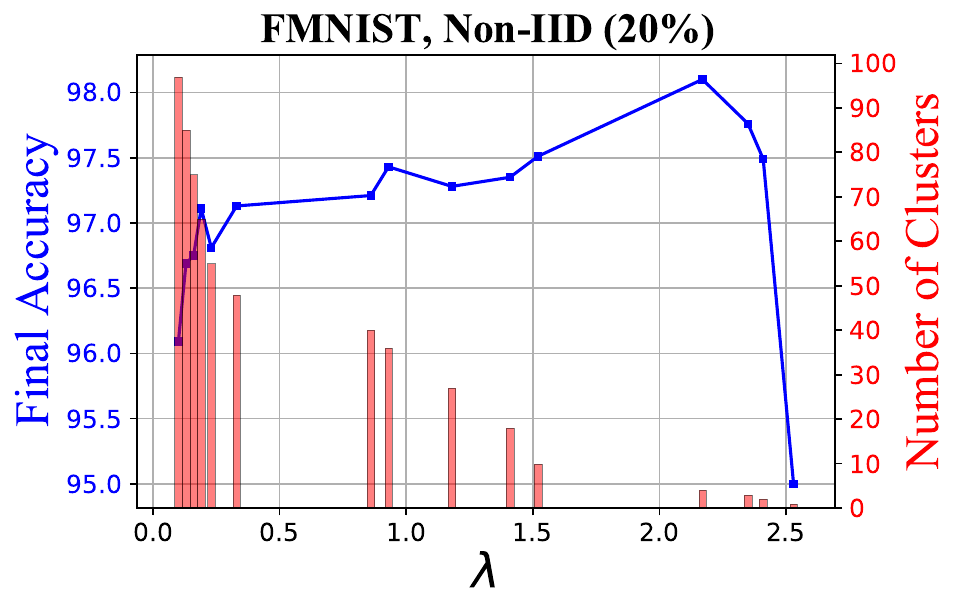}}
{\includegraphics[height=3.25cm, width=0.23\textwidth]{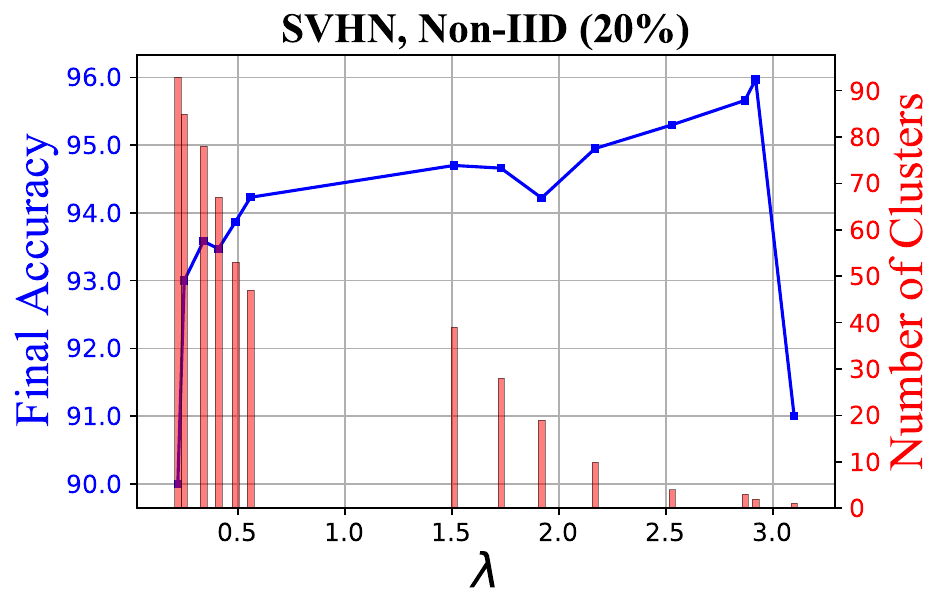}}
\caption{Test accuracy performance of FedClust versus the clustering threshold $\lambda$, and the number of suitable clusters for Non-IID label skew (20\%) on CIFAR-10/100, FMNIST, and SVHN datasets. We run each experiment to obtain each point in the plots for 200 communication rounds with local epoch and local batch size of 10, and SGD local optimizer.}
\label{figure}
\end{figure*} 

We also demonstrate the number of communication rounds needed for each baseline to reach a specified target accuracy in Table 4. The “-- --” entries in Table 4, and 5 indicate that the baseline is unable to reach the desired target accuracy. As the results show, {\em FedClust} outperforms all SOTA baseline methods. Table 5 demonstrates the required communication cost needed to reach a target test accuracy across different datasets for Non-IID label skew (30\%). The results depict that Global baseline methods are unable to reach the desired accuracy or require higher communication cost. {\em FedClust} significantly reduces the communication cost to reach the target accuracies compared to all baselines except {\em LG} (which benefits from its strategy of communicating a compact representation of raw data rather than model parameters). More specifically, with CIFAR-100, {\em IFCA} requires communication of 3495.19 Mb to achieve the desired accuracy of 50\% while {\em FedClust} only requires 1889.17 Mb. This is because the server in {\em IFCA} transmits all cluster models to participating clients in each communication round which incurs significant communication cost.  Similarly, for CIFAR-10, {\em FedClust} reduces the communication cost by (1.2 - 2.7)$\times$. 

\textbf{Impact of newcomers.}
In order to evaluate the performance of the newcomer clients personalized model, we conduct an experiment with Non-IID label skew (20\%) in which only 80 out of 100 clients are involved in a federation with 50 rounds. The remaining 20 clients are incorporated into the network after the completion of the federation and receive their corresponding cluster model from the server. The newcomer clients personalize their cluster model for only 5 epochs. The average local test accuracy of the newcomer clients is reported in Table 6. Table 6 demonstrates that {\em FedClust} has the capability to incorporate new participants to learn their personalized model with higher test accuracy.

\textbf{Trade-off between generalization and personalization.}
To address the data heterogeneity, prior works introduced a proximal term in local optimization or modified the model aggregation method on the server side to benefit from some degree of personalization \cite{li20201federated, li2019convergence}. Despite being effective, they lack the flexibility to balance between globalization and personalization. Our proposed {\em FedClust} framework can naturally navigate this trade-off. The performance of {\em FedClust} in terms of accuracy is illustrated in Fig. 4 for different values of $\lambda$, which is the clustering threshold that controls the number of clusters. The blue curve and the red bars illustrate the accuracy and number of clusters respectively for each $\lambda$. By changing the value of $\lambda$, which is determined based on the dataset, {\em FedClust} can switch from training a fully global model (1 cluster) to training fully personalized models for each client.

Fig. 4 demonstrates that increasing values of $\lambda$ lead to a decrease in the number of clusters, indicating a higher degree of globalization. {\em FedClust} groups all clients into 1 cluster when $\lambda$ is large enough and the scenario becomes similar to the {\em FedAvg} baseline (pure globalization). On the other hand, as $\lambda$ decreases, the number of clusters increases, resulting in a greater level of personalization. Each client forms individual clusters when $\lambda$ is small enough and the scenario degenerates to the Local baseline (pure personalization). The result of our experiments across all datasets demonstrates that all clients benefit from some level of globalization. For Non-IID label skew (20\%), Fig. 2 illustrates that the highest accuracy results on CIFAR-10, CIFAR-100, SVHN, and FMNIST datasets are achieved when the number of clusters are 2, 2, 2, and 4, respectively. {\em IFCA} [3] lacks this trade-off flexibility as it requires a predefined number of clusters.

\begin{table}[t]
\caption{Average local test accuracy across unseen clients on different datasets for Non-IID label skew of 20\%}
\centering
\large
\resizebox{\columnwidth}{!}{\begin{tabular}{lllll}
\toprule
\text{Method} & \text{CIFAR-10} & \text{CIFAR-100} &\text{FMNIST} &\text{SVHN}   \\
\midrule
Local& 83.39 ± 1.35 & 27.91 ± 1.09 & 94.45 ± 0.51  & 90.62 ± 0.81  \\
FedAvg& 31.72 ± 2.16 & 32.26 ± 0.48  &  78.70 ± 2.12 &  71.18 ± 3.09  \\
FedProx& 27.74 ± 2.38 & 32.74 ± 1.77  &  74.19 ± 4.17  &  73.44 ± 4.23  \\
FedNova& 31.12 ± 1.08 & 33.53 ± 0.82  & 73.76 ± 1.85 &  72.43 ± 2.91  \\
LG&81.58 ± 0.51 & 11.08 ± 0.25 & 95.66 ± 0.33 &  89.59 ± 0.90  \\
PerFedAvg& 74.65 ± 1.09 & 31.40 ± 0.36 & 92.33 ± 0.86 & 64.16 ± 1.64 \\
IFCA&85.64 ± 0.54 & 94.45 ± 0.5 & 96.63 ± 0.83 & 94.20 ± 0.15 \\
PACFL&85.80 ± 0.66 & 94.45 ± 0.5 & 97.04 ± 0.54 & 94.75 ± 0.11  \\
\textbf {FedClust}&\textbf{86.78 ± 0.67} & \textbf{97.63 ± 0.29} & \textbf{97.63 ± 0.29} & \textbf{95.19 ± 0.25 } \\
\bottomrule
\end{tabular}}
\end{table}

\textbf{Computation overhead and privacy.}
The computational overhead of {\em FedClust} is minimal compared to the {\em FedAvg} baseline algorithm. It requires performing one-shot HC after the first round. The computational complexity of {\em FedClust} is the same as {\em FedAvg}, with the additional complexity of the one-shot HC (O($N^2$)), where \textit{N} is the total number of clients. {\em FedClust} only requires each client to report selected partial weights for efficient client clustering in the first round. From the next round, {\em FedClust} leverages similar kinds of information as {\em FedAvg}. Therefore, {\em FedClust} maintains a similar level of privacy as state-of-the-art FL methods by collecting the least necessary information from clients.  

\section{Conclusion and Discussion}
In this paper, we propose a simple and effective clustered federated learning framework, {\em FedClust}, to address the data heterogeneity issue. The proposed framework aims to identify data distribution similarities among clients by exploiting the implicit relationship between the underlying data distribution and model weights. {\em FedClust} efficiently groups clients with non-IID data into an appropriate number of clusters according to the similarity among the subset of chosen weights of their locally trained models. The effectiveness of {\em FedClust} has been demonstrated through experimental evaluations over four popular datasets with a broad range of data heterogeneity scenarios.

This article includes a statistical analysis of {\em FedClust}. The convergence analysis of {\em FedClust} is left for future work. In addition, $\lambda$ is a user-defined hyperparameter which plays a crucial role in determining the number of clusters for our clustering approach. We intend to pursue a data-driven method for dynamically identifying the optimal value of $\lambda$ for each dataset in the future.

\begin{acks}
The research is supported in part by the NSF under grants OIA-2019511, OIA-2327452, 2348452, and 2315613, in part by the Louisiana BoR under LEQSF(2019-22)-RD-A-21 and LEQSF(2024-27)-RD-B-03, in part by the NSFC under 62372184, and the Sci. and Tech. Commission of Shanghai Municipality under 22DZ2229004.
\end{acks}

\bibliographystyle{ACM-Reference-Format}
\bibliography{sample-base}


\end{document}